\documentstyle[twocolumn]{article}
\newlength{\aivwidth}   
\newlength{\tmpwidth}   

\title{ On the solution of the Calogero model and its generalization to the
        case of distinguishable particles }
\author{A. Dasni\`eres de Veigy \\
        Division de Physique Th\'eorique \\
        Institut de Physique Nucl\'eaire, Orsay Fr-91406 
        \thanks{e-mail address: dasnieres@ipncls.in2p3.fr}}
\date{March 8, 1996}
\begin{document}
\sloppy
\maketitle
\begin{abstract}
The 3-body Calogero problem is solved by separation of variables for 
arbitrary exchange statistics. A numerical computation of the 4-body spectrum 
is also presented. The results display new features in comparison with 
the standard case of bosons and fermions, for instance the energies are 
not linear with the interaction parameter $\nu$ and Bethe ansatz as well as
Haldane's statistics are not verified. 
\end{abstract} \\
IPNO/TH 96-11\\
\vspace*{0.5cm}
\newcommand{\be}{\begin{equation}\label}
\newcommand{\ee}{\end{equation}}
\newcommand{\ba}{\begin{eqnarray}\label}
\newcommand{\ea}{\end{eqnarray}}
\newcommand{\ii}{{\rm i}}
\newcommand{\jj}{{\rm j}}
\newcommand{\pl}{\partial}
\newcommand{\dd}{\stackrel{\leftrightarrow}{\partial}}
\newcommand{\tra}{\, {\rm tr}\, }
\newcommand{\sign}{\, {\rm sign}\, }
\newcommand{\yt}{\mbox{\footnotesize\put(0,0){$\Box$}\put(4.0 ,0){$\Box$}
\put(8.0 ,0){$\Box$}} \hspace{0.4cm} }
\newcommand{\ydu}{\mbox{\footnotesize\put(0, 4.0){$\Box$}\put(4.0 ,4.0){$\Box$}
\put(0,0){$\Box$}} \hspace{0.4cm} }
\newcommand{\yuuu}{\mbox{\footnotesize\put(0, 8.0){$\Box$}\put(0, 4.0){$\Box$}
\put(0,0){$\Box$}} \hspace{0.4cm} }

\section{ Introduction }

 The Calogero model, that is the 1d system of $N$ bosons or $N$ fermions 
interacting by the 2-body potential $\nu(\nu-1)x^{-2}$, is of physical 
interest: it interpolates between free bosons and free fermions 
\cite{Leinaas,intermediaire}, it exhibits the fractional statistics of Haldane 
\cite{Haldane}, and its asymptotic wavefunction is in accordance with the 
Bethe hypothesis \cite{Sutherland,Marchioro}. Moreover, it belongs to a large 
class of integrable models \cite{modeles}.

 The Calogero model is known to be completely integrable for a long time,
and this fact remains true when the particles are in a harmonic well.
Calogero has solved himself the 2-body and 3-body problems by separation of 
variables. He has also shown that the many-body problem is integrable
and he has deduced its energy spectrum \cite{Calogero}. More recently, 
the many-body eigenstates have been constructed from an algebra of creation 
and annihilation operators \cite{Gambardella,Polychronakos,Brink,algebra}. 
This algebra is immediately generalizable to the case of 
Boltzmann statistics where no exchange symmetry is imposed to the wavefunction,
however the ground states and their energies are not all known in this
case.

In this paper, I discuss the asymptotic behaviour of the wavefunction when
two particles coincide and I apply it to solve the eigenvalue problem for
arbitrary statistics. 
I analytically construct the eigenvalues and eigenfunctions for $N$ bodies
in the special case $\nu=1/2$, and next for 3 bodies in the general case.
A numerical computation of the 4-body problem is also presented. 
The results obviously emcompasse the standard solution for bosons and 
fermions. However, the states of mixed statistics display a non-trivial 
framework and their energies are not linear in $\nu$. 
Moreover, the particle current does not necessarily vanish when two
particles coincide.
In conclusion, I discuss the Calogero model for distinguishable 
particles with constants $\nu_{ij}$ ,$m_i$ depending on the labels of 
the particles \cite{Furtlehener,Sen}.

\section{ The example of the two-body problem }

\subsection{ The nature of the wavefunction }

Let me focus on the relative motion of two particles of coordinates $x_1$
and $x_2$ in a harmonic well. Then the Calogero hamiltonian is 
\be{et1} 
H=-\partial^2+{1\over4}\omega^2x^2+{\nu(\nu-1)\over x^2},
\ee
with $x=x_1-x_2$, $\hbar=1$ and $m=1$. One will assume $0<\nu<1$ for 
simplicity, the limits $\nu=0$ and $\nu=1$ corresponding to the usual
harmonic oscillator problem. The hamiltonian acts on a domain of 
definition 
\be{et2}
D_H\subset\{\psi,\ \psi \in L^2,H\psi\in L^2,H^+=H\}
\ee 
which warrants finite matrix elements, real eigenvalues and the 
orthogonality between states of different energies. There remains to
understand the nature of the wavefunctions $\psi\in D_H$.

At ordinary points $x\ne0$, the wavefunction and its derivative may be
assumed continuous as usual.
Around the singular point $x=0$, one verifies that the wavefunction must 
have an algebraic behaviour as 
\be{et3}
\psi\sim x^\nu \quad {\rm or} \quad \psi\sim x^{1-\nu},
\ee
otherwise its image $H\psi$ is not a square integrable function.
Note that the situations $\nu<0$ and $\nu>1$ will not be considered 
in this paper because the wavefunction is then divergent in general.

The hamiltonian is self-adjoint when the boundary term
\be{et4}
\int^x dx \ (\bar\psi_1\partial^2\psi_2 -\psi_2\partial^2\bar\psi_1)=
\bar\psi_1 \dd \psi_2 \ (x)
\ee
is continuous at any point and vanishes at infinity for all pair of 
wavefunctions $\psi_1,\psi_2$. Notice that ${1\over2\ii} \, \bar\psi\dd \psi$ 
is merely the particle current. It is equivalent to impose the previous 
condition on the boundary term for all pair of basic states $\psi_1,\psi_2$, 
or the same condition on the particle current for all wavefunctions $\psi$. 
This condition is satisfied at ordinary points and at infinity because 
the wavefunction and its first derivative are continous at these points 
and vanishe at infinity. At the singular point $x=0$, one can always set
\be{et5} 
\psi= |x|^\nu f(x)+ x|x|^{-\nu} g(x)
\ee
and then one has 
\ba{et6}
\bar\psi_1\dd \psi_2 &=& (1-2\nu)(\bar f_1g_2-\bar g_1f_2)
+x(\bar f_1\dd g_2+\bar g_1\dd f_2) \nonumber\\
& & +|x|^{2\nu}(\bar f_1\dd f_2) +x^2|x|^{-2\nu}(\bar g_1\dd g_2).
\ea
For $0<\nu<1$ and $\nu\ne1/2$, the continuity of the particle current is 
ensured at $x=0$ if $f$ and $g$ are continuous up to a given phase shift 
at this point and if $\partial f$ and $\partial g$ are finite at this point. 
The case $\nu=1/2$ is particular because the continuity of $f$ and $g$ 
can then be relaxed at $x=0$.\footnote{
Note that the regular combination $|x|^{1/2} f(x) +x|x|^{-1/2} f(x)= 2
\delta_{x>0} \sqrt{|x|} f(x)$ leads to a discontinuity as $2\delta_{x>0} 
f(x)$.}
The choice of the overall phase shift is essentially arbitrary. 
This phase shift can be removed by adding a constant phase depending on 
the ordering of the particles, $x>0$ or $x<0$. 
As a result, one can demand that $f$ and $g$ are exactly continuous at $x=0$. 
An alternative choice will be presented below in relation with the notion 
of intermediate statistics.

The differential equation theory shows that $f$ and $g$ are entire series
in $x$ \cite{Wang}. These functions and their derivatives are so simultaneously 
continuous at $x=0$. In the limit $\nu=0$, the problem is reduced to that 
of a pure harmonic oscillator, the wavefunction and its derivative are 
continuous as it should and the usual harmonic eigenstates are necessarily 
recovered. 

Note that both the hamiltonian and the boundary conditions are invariant 
under the mirror transformation $\psi(x,\nu)\to \sign(x)\, \psi(x,1-\nu)$. 
Introducing the odd factor $\sign(x)$, the mirror symmetry inverts the 
exchange properties of $\psi$, that is, it exchanges Bose and Fermi 
statistics.

\subsection{ An anyon gauge }

In analogy with the 2d anyon model \cite{anyons}, it is possible to define 
a new gauge by the gauge transformation
\be{7} 
\psi'=e^{\ii\nu\theta}\psi,
\ee
with $\theta=\arg(x)$ i.e. $\theta=0$ if $x\ge0$ and $\theta=\pi$ if $x<0$.
Let me name it the anyon gauge. In this new gauge, the hamiltonian is the
same but the boundary condition at $x=0$ becomes
\be{8}
\psi'=x^\nu f(x)+\bar x^{1-\nu} g(x)
\ee
with $f$ and $g$ continuous. The last equation is well defined if 
a cut is introduced in the complex plane $z=x+\ii y$ along the axis $x<0$, 
and if, on this axis, $x$ is defined by the analytic continuation 
$x=\lim_{ \epsilon\to0^+}|x| e^{\ii(\pi-\epsilon)}$. 

In the anyon gauge, the wavefunction is multivalued when two identical
particle interchanges are successively effected and, thus, the wavefunction
is neither symmetric nor antisymmetric but it has complicated exchange
symmetries. However the limits $\nu=0$ and $\nu=1$ reproduce monovalued
wavefunctions: the gauge transformation is the identity in the case $\nu=0$ 
and it multiplies the wavefunction by the odd factor $\sign(x)$ in the case 
$\nu=1$. As a result, if the wavefunction $\psi$ satisfies Bose statistics, 
the wavefunction $\psi'$ will be bosonic at $\nu=0$ and fermionic at $\nu=1$, 
and vis versa if $\psi$ satisfies Fermi statistics.

\subsection{ Eigenvalues and eigenfunctions }

The eigenstates are obtained in a standard way. Let me choose the regular 
gauge characterized by the boundary condition (\ref{et5}). Thus, the bosonic 
eigenstates are
\be{5}
\psi_n =|x|^\nu L_n^{\nu-{1\over2}} ({1\over2}\omega x^2) e^{-{1\over4}
\omega x^2}, \quad E_n=(2n+{1\over2}+\nu)\omega,
\ee
and the fermionic eigenstates are
\be{6}
\psi_n =x|x|^{-\nu} L_n^{{1\over2}-\nu} ({1\over2}\omega x^2) e^{-{1\over4}
\omega x^2}, \quad E_n=(2n+{3\over2}-\nu)\omega,
\ee
where $n$ is a non-negative integer. The whole basis is clearly 
orthogonal and complete in $L^2$. One remarks the mirror symmetry between
the Bose and the Fermi states. Using the identities $H_{2n}(z)=
(-)^n2^{2n}n!L_n^{-1/2}(z^2)$ and $H_{2n+1}(z) =(-)^n2^{2n+1}n!z
L_n^{1/2}(z^2)$ where $H_n$ is the Hermite polynomial, the harmonic 
oscillator solution is reproduced in the limits $\nu=0$ and $\nu=1$, 
except that there is a phase shift at $x=0$ in the last case.

The eigenstates of the anyon gauge are merely deduced from the changes 
$|x|^\nu\to x^\nu$ and $|x|^{-\nu} \to \bar x^{-\nu}$. They interpolate 
between Bose and Fermi statistics as it should.

\section{ Exact results for the many-body problem }

\subsection{ The model }

The $N$-body hamiltonian is
\be{m1}
H=\sum_{i=1}^N \big(-{1\over2}\partial_i^2 +{1\over2}\omega^2 x_i^2\big)
+\sum_{i,j<} {\nu(\nu-1)\over x_{ij}^2}
\ee
with $x_{ij}=x_i-x_j$. The wavefunction $\psi$ has to be sought in an 
irreducible representation of the permutation group $S_N$, usually bosonic 
or fermionic.
Let me still assume $0<\nu<1$, the limits $\nu=0$ and $\nu=1$ reproducing
the standard harmonic oscillator problem. The domain of definition of the 
hamiltonian is chosen as for the 2-body case : the wavefunction and its 
derivative are continuous except at $x_i=x_j$ where
\be{m2}
\psi= |x_{ij}|^\nu f +x_{ij}|x_{ij}|^{-\nu} g 
\ee
with $f$ and $g$ locally continuous, $\partial_{ij}f$ and $\partial_{ij}g$
finite ($\partial_{ij}$ is a shorted notation for $\partial_i-\partial_j$).
The case $\nu=1/2$ is special: the continuity of $f$ and $g$ is no longer
necessary and the boundary condition at $x_i=x_j$ is reduced to 
\be{m3} 
\psi= \sqrt{|x_{ij}|} \; h
\ee
with $\partial_{ij}h$ finite.

As for the 2-body case, we can define an anyon gauge by 
\be{m4}
\psi'=e^{\ii\nu\sum_{i,j<}\theta_{ij}} \psi
\ee
where $\theta_{ij}=\arg(x_{ij})$. The hamiltonian is unaltered, but the 
boundary condition at $x_i=x_j$ is changed into
\be{m5}
\psi'= x_{ij}^\nu f+ \bar x_{ij}^{1-\nu} g.
\ee
The standard solution of the harmonic oscillator is then exactly reproduced 
at $\nu=0$ and $\nu=1$. However, at $\nu=1$, the anyon gauge (\ref{m4}) 
introduces an antisymmetric factor $\sign(\prod_{i,j<} x_{ij})$ which 
inverts the exchange properties of the wavefunction: 
if $\psi$ belongs to an irreducible representation of $S_N$, then its image 
$\psi'$ belongs to the conjugate representation since the antisymmetric factor 
exchanges lines (symmetrisors) and columns (antisymmetrisors) commuting
with the Young diagramm (projector).
As a result, when $\nu$ goes over $[0,1]$, the particles interpolate between 
bosons and fermions, or more generally, between an irreducible representation 
and the conjugate representation of the permutation group $S_N$. 

The mirror symmetry reads $\psi'(x_i,\nu)\to \psi'(\bar x_i,1-\nu)$ in the
anyon gauge and $\psi(x_i,\nu)\to \sign(\prod_{i,j<} x_{ij}) \psi(x_i,1-\nu)$ 
in the regular gauge. In the regular gauge, the mirror transformation inverts 
the exchange properties of the wavefunction introducing an antisymmetric factor.

Working with an unperturbed basis, the matrix elements of the interaction 
are divergent. In analogy with the 2d anyon model, the non-unitary 
transformation
\be{m6}
\psi'= \prod_{i,j<} x_{ij}^\nu \; \widetilde\psi \quad \iff \quad
\psi= \prod_{i,j<} |x_{ij}|^\nu \; \widetilde\psi 
\ee
(or its mirror symmetric) leads to a hamiltonian 
\be{m7}
\widetilde H=\sum_{i=1}^N \big(-{1\over2}\partial_i^2 +{1\over2}\omega^2 
x_i^2 \big) -\sum_{i,j<}{\nu\over x_{ij}}(\partial_i-\partial_j)
\ee
(or respectively its mirror symmetric) whose matrix elements are well
defined with the principal value regularization. It is especially adapted 
to the perturbative and numerical analyses. Albeit it is not formally 
hermitian, it is so between states of the same unperturbed energy.

\subsection{ Separation of two variables }

One defines the center-of-mass coordinate $X={1\over N}\sum_i x_i$ and
a radial coordinate for the relative motion $r^2={1\over N} \sum_{i,j<}
x_{ij}^2$. The remaining coordinates constitute a system of angles denoted 
by $\Omega$. In these variables, the hamiltonian reads 
\be{st1}
H= -{1\over2N}\partial_X^2 +{1\over2}N\omega^2X^2
-{1\over2}\partial_r^2 -{N\! -\! 2\over2r}\partial_r +{\Lambda\over2r^2}
+{1\over2}\omega^2r^2.
\ee
where $\Lambda$ is an angular operator. The eigenfunctions may be separated 
as $\psi= \Xi(X) R(r) Y(\Omega)$. Assuming that the eigenvalue equation 
$\Lambda Y=\lambda Y$ is satisfied, the normalizable solutions for the 
center of mass 
\be{st2}
\Xi_\kappa (X) =H_\kappa(\sqrt{N\omega}X) e^{-{1\over2}N\omega X^2}, \quad
\ee
and for the radial part
\be{st3}
R_n(r) =r^{\sqrt{\lambda+ \big( {N-3\over2} \big)^2 } -{N-3\over2}}
L_n^{\sqrt{\lambda+ \big( {N-3\over2} \big)^2 }} (\omega r^2) 
e^{-{1\over2}\omega r^2}, 
\ee
are standard. The total energy is then given by
\be{st4}
E_{\kappa,n,\lambda} =\left( \kappa+{1\over2} +2n+1+ \sqrt{\lambda+ 
\Big({N-3\over2}\Big)^2} \; \right) \omega.
\ee
$\kappa$ and $n$ are non-negative integers.

\subsection{ Raising and lowering operators }

This section is a brief account of some recent progress. 
The main idea consists to construct a covariante derivative using the
operator $P_{ij}$ which exchanges the particles $i$ and $j$ 
\cite{Polychronakos,Brink},
\be{so1}
D_i= \partial_i -\sum_{j\ne i} {\nu\over x_{ij}} P_{ij}.
\ee
The adjoint of $D_i$ is merely $-D_i$, for $P_{ij}$ and $x_{ij}^{-1}$ are 
anticommuting. The commutation rules are $[D_i,D_j]=0$ and 
\be{so2}
[D_i,x_j]= \delta_{ij} (1+\sum_{k=1}^N\nu P_{ik}) -\nu P_{ij}.
\ee
One next defines the hamiltonian
\be{so3}
{\cal H}= {1\over2} \sum_{i=1}^N \big( a_i^+a_i^- +a_i^-a_i^+ \big)\omega,
\ee
in terms of creation and annihilation operators,
\be{so4}
a_i^{\pm}= \sqrt{\omega\over2}x_i \mp {D_i\over\sqrt{2\omega}}, \quad
[{\cal H},a_i^\pm]=\pm\omega a_i^\pm. 
\ee
These operators are mutually adjoints. 

As regards its action on bosonic wavefunctions, the hamiltonian ${\cal H}$
is found to be identical with the Calogero one. As a result, the bosonic
eigenstates of the Calogero model are easily obtained. The Bose groundstate 
\be{so5}
\prod_{i,j<}|x_{ij}|^\nu \; e^{-{1\over2}\omega \sum x_i^2}
\ee
is annihilated by the $a_i^-$'s, its energy is ${N\over2}\omega+ {N(N-1)\over2}
\omega\nu$, and the excitements follow from the successive actions of the 
$a_i^+$'s on the groundstate. However, these excitements have to be symmetrized 
under particle exchanges. To avoid this symmetrization, it is desirable 
to define a set of operators preserving the exchange symmetries of 
the wavefunction. Separating the center-of-mass and the relative excitements,
such a set of Raising and lowering operators is given by 
\cite{Gambardella}
\be{so6}
{\cal A}_1^\pm= \sum_{i=1}^N a_i^\pm, \quad 
{\cal A}_{k\ge2}^\pm= \sum_{i=1}^N (a_i^\pm -{1\over N}{\cal A}_1^\pm)^k, 
\ee
The set is complete if $k=1,2,\ldots,N$. For a discussion of the algebra
associated with these operators, see \cite{algebra}. The main commutation 
rules are
\be{so7}
\quad [{\cal H},{\cal A}_k^\pm]= 
\pm \, k \, \omega {\cal A}_k^\pm.
\ee
In this way, the bosonic excitements are directly obtained as 
\be{so8}
\psi_{n_k}= \prod_{k=1}^N ({\cal A}^+_k)^{n_k} 
\prod_{i,j<}|x_{ij}|^\nu  \; e^{-{1\over2}\omega \sum x_i^2}, 
\ee
and the total energies are
\be{so9}
E_{n_k}= \sum_{k=1}^N k \, n_k \, \omega +{N\over2}\omega 
+{N(N-1)\over2}\omega\nu,
\ee 
where the $n_k$'s are non-negative integers. Here, the net effect of the 
raising operators is to multiply the groundstate by a polynomial.
The particle current (\ref{et6}) vanishes at coinciding points $x_i=x_j$ 
because $g=0$ and thus there is no particle exchange.

As regards its action on an arbitrary function, the hamiltonian ${\cal H}$ 
is a non-local operator effecting among others some interchanges on the 
variables of the function. It is necessarily different from the Calogero 
hamiltonian $H$. However, $H$ is the only local operator which coincides 
with ${\cal H}$ acting on symmetric functions. Let me define $A_k^\pm$ 
as the only local operator which coincides with ${\cal A}_k^\pm$ acting 
on symmetric functions. 
The explicit form of $A_k^\pm$ can be obtained by expanding ${\cal A}_k^\pm$
in power series of $P_{ij}$, commuting the $P_{ij}$'s in first position and 
then replacing $P_{ij}$ by unity. One finds in this way
\be{so10}
A_1^\pm= {\cal A}_1^+= N\sqrt{\omega\over2}\, X\mp 
{\partial_X\over\sqrt{2\omega}}, 
\ee
\be{so11}
A_2^\pm=-{H_{\rm rel.}\over\omega} +\omega r^2\mp r\partial_r\mp{N+1\over2},
\ee
\ba{so12}
A_3^\pm &=& \sum_{i=1}^N 
(\sqrt{\omega\over2}x_i\mp{\partial_i\over\sqrt{2\omega}}-{1\over N}A_1^\pm)^3
\\ & & -3\sum_{i,j\ne} {\nu(\nu-1)\over 2\omega x_{ij}^2} (\sqrt{\omega\over2} 
x_i \mp{\partial_i\over\sqrt{2\omega}} -{1\over N}A_1^\pm), \nonumber
\ea
et caetera \cite{Gambardella}. One can directly check that $A_1^+$ and
$A_2^+$ respectively originate the center-of-mass excitements (\ref{st2}) 
and the radial excitements (\ref{st3}). 

By definition, the operator $A_k^\pm$ does not alter the exchange symmetries 
of the wavefunctions. Moreover, in a Hilbert space of symmetric functions, 
the commutation rule (\ref{so7}) can be immediately translated into
\be{so13}
[H,A_k^\pm]= \pm\, k\, \omega A_k^\pm
\ee
Now, this relation involves only local operators so that it is also 
satisfied independently of the exchange symmetries of the wavefunctions. 
One concludes that the $A_k^\pm$'s are Raising and lowering operators 
for arbitrary exchange statistics. In fact, it remains to prove that
the action of these operators on an eigenfunction having the correct
behaviour at coinciding points gives a new eigenfunction having also this 
behaviour. Since the operators involve derivatives, it suffices to
assume that the successive derivatives of $f$ and $g$ in (\ref{m2}) are all
continuous at $x_i=x_j$. Every eigenfunction discussed in this paper 
confirm this hypothesis.

In conclusion, the ground states of the Calogero model are annihilated by 
the $A_k^-$'s and the excitements are obtained by the action of $\prod_{k=1}^N 
(A_k^+)^{n_k}$ on these ground states. The energy spectrum is linear in the 
quantum numbers $n_k$. According to the permutation theory, a complete 
eigenstate basis will be obtained if there are $N!$ independent ground states:
one in the Bose representation, another in the Fermi representation, and $d$ 
ground states in each irreducible representation of degree $d$ of $S_N$.

\subsection{ Eigenstates and eigenvalues for bosons and fermions }

The Bose eigenstates are discussed in the preceding paragraph.
The Fermi eigenstates can be deduced from the Bose ones by using the mirror 
symmetry.

\subsection{ Eigenstates and eigenvalues at $\nu=1/2$ }

At $\nu=1/2$, the behaviour of the wavefunction when two particles coincide
is dictated by (\ref{m3}) where the continuity of $h$ is no longer required.
It is possible to construct an eigenstate $\psi_{p,n_k}$ 
equals to the Bose eigenstate $\psi_{n_k}$ in the sector 
$x_{p_1} >x_{p_2} >\ldots >x_{p_N}$ and sets to zero in the other sectors. 
The states which are non-zero on different sectors are obviously linearly 
independent. Hence one has the eigenstate basis
\be{eea1}
\psi_{p,n_k}=\delta_{x_{p_1}>x_{p_2}>\ldots >x_{p_N}}
\prod_{k=1}^N ({\cal A}_k^+)^{n_k} \; \prod_{i,j<} \sqrt{|x_{ij}|} \; 
e^{-{1\over2}\omega\sum x_i^2}
\ee
with the energy spectrum
\be{eea2}
E_{p,n_k}= \sum_{k=1}^N k\, n_k\, \omega +{N(N+1)\over4}\omega.
\ee
The quantum numbers are the permutations $p$ and the non-negative integers
$n_k$.

Note that the energies are completely degenerated with respect to particle 
exchanges because the action of the permutations on any eigenstate gives $N!$ 
independent eigenstates. This last property implies the completness of the 
basis (\ref{eea1}) in $L^2$ provided that the subspace of the bosonic 
eigenstates is complete.

Physically, at $\nu=1/2$, the particle current always vanishes when two 
particles coincide, for all statistics. There is no exchange of particles 
and no change of the ordering of the particles as in classical mechanics
\cite{intermediaire}. 
In this situation, the coupling strength $\nu(\nu-1)$ is at its 
minimal value $-1/4$. Indeed, for a smaller strength, $\nu$ is a complex
number, the energies are also complex and this indicates that the particle 
system collapses.

\section{ General solution for three particles }

\subsection{ System of coordinates }

Introducing the following Jacobi coordinates,
\ba{sc1}
X &=& {1\over3} (x_1+x_2+x_3),  \nonumber\\
y &=& {1\over\sqrt{2}} (x_1-x_2), \\
x &=& {1\over\sqrt{6}} (x_1+x_2-2x_3), \nonumber
\ea
the relative coordinates $x,y$ are expressed as functions of the standard 
polar coordinates $r,\phi$ \cite{Calogero}. The radial coordinate also 
satisfies the alternative definition $r^2={1\over3} \sum_{i,j<}x_{ij}^2$. 
This system of coordinates is related to a discrete Fourier transformation 
of the coordinates on the line,
\be{sc2}
r e^{\ii\phi}= -\sqrt{2\over3} \; (\jj^2 x_1+\jj x_2+ x_3)
\ee
where $\jj=e^{ \ii2\pi/3}$ is a cube root of unity, so that its exchange 
properties are very simple. The cyclic exchange $(x_1,x_2,x_3)\to(x_2,x_3,
x_1)$ is identical with $\phi \to\phi+{2\over3}\pi$ and the interchange 
$x_1\leftrightarrow x_2$ is identical with $\phi\to -\phi$. Note that these 
two permutations generate the whole symmetric group $S_3$.

One easily obtains the interparticle coordinates
\ba{sc3}  
&x_{12}=\sqrt{2}\, r \sin\phi, \quad
x_{23}=\sqrt{2}\, r \sin(\phi+{2\pi\over3}),& \\
&x_{31}=\sqrt{2}\, r \sin(\phi+{4\pi\over3}).& \nonumber 
\ea From these relations, one deduces that the value of $\phi$ corresponds 
to the ordering of the three particles as follows,
\ba{sc4}
\ \ 0<\phi<{\pi\over3}         &\iff&  x_1>x_2>x_3, \nonumber\\
{\pi\over3}<\phi<2{\pi\over3}  &\iff&  x_1>x_3>x_2, \nonumber\\
\ 2{\pi\over3}<\phi<\pi        &\iff&  x_3>x_1>x_2, \nonumber\\
\ \pi<\phi<4{\pi\over3}        &\iff&  x_3>x_2>x_1, \\
4{\pi\over3}<\phi<5{\pi\over3} &\iff&  x_2>x_3>x_1, \nonumber\\
5{\pi\over3}<\phi<2\pi         &\iff&  x_2>x_1>x_3, \nonumber
\ea
At a bound $\phi=n\pi/3$ with $n$ integer, two particles coincide.
On the other hand, note the useful identities
\be{sc5} 
\sin\phi \; \sin(\phi+{2\pi\over3}) \; \sin(\phi+{4\pi\over3})
=-{1\over4}\sin3\phi  
\ee
\be{sc6} 
{1\over\sin^2\phi}+ {1\over\sin^2(\phi+{2\pi\over3})}+
{1\over\sin^2(\phi+{4\pi\over3})}= {9\over\sin^23\phi}.  
\ee

\subsection{ Separation of variables }

In the case of $N=3$ particles, the wavefunction is completely separable as 
$\psi= \Xi(X) R(r) \Phi(\phi)$ according to (\ref{st2}) and (\ref{st3}). 
Using (\ref{sc6}), the angular eigenvalue equation $\Lambda\Phi =\lambda\Phi$ 
reads
\be{sv1}
-\partial_\phi^2\Phi +{9\nu(\nu-1)\over\sin^2{3\phi}}\Phi 
=\lambda\Phi.
\ee

Since the particles are identical, the angular equation can be solved
in the irreducible representations of $S_3$,
\be{sv2}
S_3=\yt\; +2\ydu+\yuuu \, .
\ee
The Bose and Fermi representations are of degree 1, and the two mixed 
representations are equivalent and each of degree 2. Let me choose to 
simultaneously diagonalize the hamiltonian and the cyclic permutations.
The eigenvalues of a cyclic permutation are $1,\jj,\jj^2$ because the
cube of a cyclic permutation is merely the identity permutation. 

The bosonic (fermionic) states are invariant under the cyclic exchanges
and they are even (odd) under the interchange of two particles,
\be{sv3}
\yt\; : \quad \Phi(\phi+{2\pi\over3})=\Phi(\phi)=\Phi(-\phi),
\ee
\be{sv4}
\yuuu : \quad \Phi(\phi+{2\pi\over3})=\Phi(\phi)=-\Phi(-\phi),
\ee

The remaining states belong to the mixed representations,
\be{sv5}
\ydu : \quad \Phi(\phi+{2\pi\over3})=r \Phi(\phi), \quad 
r=\jj,\jj^2.
\ee
The states with $r=\jj$ and the ones with $r=\jj^2$ are in correspondence
under the complex conjugation, and also under the interchange of two
particles (doing $\Phi(\phi)\to\Phi(-\phi)$ and $\phi\to -\phi-{2\over3}
\pi$) in conformity with the degree 2 of the mixed representations.
Their energy spectrums are identical.

\subsection{ Angular quantization }

The differential equation (\ref{sv1}) may be transformed into the hypergeometric
equation by the redefinition $\Phi= |\sin3\phi|^\nu \; \widetilde\Phi$ and the 
substitution $z=\cos^2 3\phi$. It follows that the angular eigenvalue equation 
only has two linearly independent solutions,
\be{aq1}
\Phi_1(\phi)= |\sin3\phi|^\nu F({\mu+\nu\over2}, 
{\nu-\mu\over2}; {1\over2}; \cos^2{3\phi}),
\ee
\be{aq2}
\Phi_2(\phi)= |\sin3\phi|^\nu \cos3\phi 
\ F({1+\mu+\nu\over2}, {1-\mu+\nu\over2}; {3\over2}; \cos^2{3\phi}),
\ee
where $\mu={1\over3}\sqrt{\lambda}$. According to the theory of differential
equations, only the singularities of the differential equation could be
singularities of its solutions. These singularities occur at $\phi=n\pi/3$,
when two particles coincide, and they are of algebraic type, 
namely\footnote{
by using the relation
$F(a,b;c;z)= {\Gamma(c)\Gamma(c-a-b) \over\Gamma(c-a)\Gamma(c-b)}
F(a,b;a+b-c+1;1-z) +(1-z)^{c-a-b} {\Gamma(c)\Gamma(a+b-c) 
\over\Gamma(a)\Gamma(b)} F(c-a,c-b;c-a-b+1;1-z)$, and also the relation
$F(a,b;c;z)= (1-z)^{c-a-b} F(c-a,c-b;c;z)$ in the case of $\Phi_2$.
},
\ba{aq3}
\Phi_1(\phi) &=& \vspace*{-0.8cm} \gamma_{11} |\sin3\phi|^\nu F( {{\mu+\nu}
\over{2}}, {{\nu-\mu}\over{2}}; {{1+2\nu}\over{2}}; \sin^2{3\phi} ) \nonumber\\
& & +\gamma_{12} |\sin3\phi|^{1-\nu} \\
& & \times F( {{1-\mu-\nu}\over{2}}, 
{{1+\mu-\nu}\over{2}}; {{3-2\nu}\over{2}}; \sin^2{3\phi} ), \nonumber
\ea
\ba{aq4}
\Phi_2(\phi) &=& \vspace*{-0.8cm} \gamma_{21}|\sin3\phi|^\nu\sign(\cos3\phi) 
\nonumber \\ 
& & \times F( {{\mu+\nu}\over{2}}, {{\nu-\mu}\over{2}}; {{1+2\nu}\over{2}}; 
\sin^2{3\phi} ) \\
& & +\gamma_{22} |\sin3\phi|^{1-\nu} \sign(\cos3\phi) \nonumber \\
& & \times F( {{1-\mu-\nu}\over{2}}, 
{{1+\mu-\nu}\over{2}}; {{3-2\nu}\over{2}}; \sin^2{3\phi} ), \nonumber
\ea
with
\ba{aq5}
\gamma_{11} ={\Gamma({1\over2}) \Gamma({1-2\nu\over2}) \over
\Gamma({1+\mu-\nu\over2}) \Gamma({1-\mu-\nu\over2}) } &,& {\displaystyle
\gamma_{12} ={\Gamma({1\over2}) \Gamma({2\nu-1\over2}) \over
\Gamma({\mu+\nu\over2}) \Gamma({\nu-\mu\over2})}       },
\nonumber\\
\gamma_{21} ={\Gamma({3\over2}) \Gamma({1-2\nu\over2}) \over
\Gamma({2+\mu-\nu\over2}) \Gamma({2-\mu-\nu\over2}) } &,& {\displaystyle
\gamma_{22} ={\Gamma({3\over2}) \Gamma({2\nu-1\over2}) \over
\Gamma({1+\mu+\nu\over2}) \Gamma({1-\mu+\nu\over2})}    }, \nonumber
\ea
where $\Gamma$ is the gamma function. 

At an ordinary point, the eigenfunction is a combination of these 
two independent solutions,
\be{aq6}
\Phi(\phi)= \alpha_n\Phi_1(\phi) +\beta_n\Phi_2(\phi),
\quad \phi\in]n{\pi\over3},(n+1){\pi\over3}[.
\ee
At a singular point $\phi=n\pi/3$ two particles coincide so that the 
asymptotic behaviour of the wave function must be 
\be{aq7}
\Phi(\phi)= |\sin3\phi|^\nu f(\phi) +\sin3\phi \; 
|\sin3\phi|^{-\nu} g(\phi)
\ee
with $f$ and $g$ continuous, in conformity with (\ref{m2}), (\ref{sc3}) 
and (\ref{sc5}). The general form of the eigenfunction (\ref{aq6}) reproduces 
the algebraic singularities $|\sin3\phi|^\nu$ and $\sin3\phi\; |\sin3 
\phi|^{-\nu}$, and the functions $f$ and $g$ in factor are continuous 
at $\phi=n\pi/3$ when one has
\be{aq8}
\pmatrix{ \gamma_{11} &(-)^n\gamma_{21} \cr \gamma_{12} &(-)^n\gamma_{22} \cr}
\pmatrix{ \alpha_{n-1} \cr \beta_{n-1} \cr}=
\pmatrix{ \gamma_{11} &(-)^n\gamma_{21} \cr -\gamma_{12} &-(-)^n\gamma_{22} \cr}
\pmatrix{ \alpha_n \cr \beta_n \cr}.
\ee
Note that the functions $f$ and $g$ are then proportional to hypergeometric 
series: their successive derivatives are also continuous at $\phi=n\pi/3$.
The recurrence relation (\ref{aq8}) may be rewritten as 
\be{aq9}
\pmatrix{ \alpha_n \cr \beta_n \cr} =T_n
\pmatrix{ \alpha_{n-1} \cr \beta_{n-1} \cr}
\ee
with the transfer matrix
\ba{aq10}
T_n &=& {1\over \gamma_{11}\gamma_{22} -\gamma_{12}\gamma_{21}} \\
& & \times\pmatrix{ \gamma_{11}\gamma_{22}+\gamma_{12}\gamma_{21} 
               &2(-)^n\gamma_{21}\gamma_{22} \cr 
               -2(-)^n\gamma_{11}\gamma_{12} 
               &-\gamma_{11}\gamma_{22}-\gamma_{12}\gamma_{21} \cr}.
\nonumber
\ea
For a given $(\alpha_0,\beta_0)$, this recurrence relation determines 
$(\alpha_1,\beta_1),\ldots, (\alpha_5,\beta_5)$ and thus the wave function is 
obtained on the period $[0,2\pi]$.
However, only the wave functions with period $2\pi$ are physically acceptable, 
and such functions are only for quantized values of $\lambda$.

Let me perform the quantization taking advantage of the classification into
representations of $S_3$. The periodicity requirement is then replaced by
$\Phi(\phi+{2\over3}\pi)=r\Phi(\phi)$ with $r=1$ for bosons and 
fermions, and $r=\jj,\jj^2$ for the mixed representation. This may
be expressed as
\be{aq11}
T_2T_1 \pmatrix{ \alpha_0 \cr \beta_0 \cr} = r 
\pmatrix{ \alpha_0 \cr \beta_0 \cr}
\ee
and has a solution when $\det(T_2T_1-r)=0$. This last equation is 
responsible for the quantization of $\lambda$, it may be expanded as 
\be{aq12} 
\det T_2T_1 -r\, \tra T_2T_1 +r^2=0.
\ee
Now, one has $\det T_n=-1$ and 
\be{aq13} 
\tra T_2T_1 =4\left( \gamma_{11}\gamma_{22}+\gamma_{12}\gamma_{21}
\over \gamma_{11}\gamma_{22}-\gamma_{12}\gamma_{21} \right)^2 -2.
\ee
Applying the product formula for the gamma function,
\be{aq14}
\gamma_{11}\gamma_{22} +\gamma_{12}\gamma_{21} =-{1\over1-2\nu}
{\cos\mu\pi \over\cos\nu\pi},
\ee
\be{aq15}
\gamma_{11}\gamma_{22} -\gamma_{12}\gamma_{21} =-{1\over1-2\nu}.
\ee
Finally, the angular eigenvalue $\lambda$ is discretized according to
\be{aq16}
\cos^2\mu\pi ={(1+r)^2\over4r} \cos^2\nu\pi, \quad 
\sqrt{\lambda}=3\mu\ge0.
\ee
The solution is immediate by now. Fig.1 displays the discrete values of 
$\sqrt{\lambda}$ for the 3! ground states. One clearly sees the mirror 
symmetry and, at $\nu=1/2$, the crossing associated with the degenerate 
solution (\ref{eea2}).

The Bose representation is characterized by $r=1$ and $\mu=\ell+\nu$ where 
$\ell$ is a non-negative integer. 
One has then $\gamma_{12}=0$, $\beta_n=0$, $\alpha_n=\alpha_0$ if $\ell$ 
is even; $\gamma_{22}=0$, $\alpha_n=0$, $\beta_n=\beta_0$ if $\ell$ is odd. 
In each case, the remaining hypergeometric function degenerates into 
a Gegenbauer polynomial $C_\ell^\nu$, hence the Bose eigenstate basis
\be{aq17}
\Phi_\ell(\phi)= |\sin3\phi|^\nu C_\ell^\nu(\cos3\phi), \quad
\sqrt{\lambda_\ell}=3\ell+3\nu.
\ee
At $\nu=1$, the free states of Fermi statistics are reproduced but with
some phase shifts at $\phi=n\pi/3$.

The Fermi representation is characterized by $r=1$ and $\mu=\ell+1-\nu$ where 
$\ell$ is a non-negative integer. One has then $\gamma_{11}=0$, $\beta_n=0$, 
$\alpha_n=\alpha_0$ if $\ell$ is even; $\gamma_{21}=0$, $\alpha_n=0$, $\beta_n
=\beta_0$ if $\ell$ is odd. One recovers the Fermi eigenstate basis obtained 
by Calogero,
\be{aq18}
\Phi_\ell(\phi)= \sin3\phi |\sin3\phi|^{-\nu} C_\ell^{1-\nu}
(\cos3\phi), \quad \sqrt{\lambda_\ell}=3\ell+3-3\nu. 
\ee
The fermi eigenstates correspond to the Bose ones under the mirror symmetry.

At least, the mixed representation involves the remaining two possibilities
$r=\jj$ and $r=\jj^2$. The eigenvalues are solutions of the same equation 
$\cos^2{\mu\pi}= {1\over2} \cos^2{\nu\pi}$ in the two cases so that they 
are two-fold degenerate. One obtains
\be{aq19}
\sqrt{\lambda}= 3\ell+{3\over2} \mp{3\over\pi}
\arcsin\left({\cos\nu\pi \over 2}\right),
\ee
where $\ell$ is a non-negative integer. These eigenvalues reproduce the 
non-linear levels in Fig.1, the sign $-$ is responsible for the increasing 
levels and the sign $+$ is for the corresponding ones under the mirror 
symmetry. The eigenstates directly follow from (\ref{aq9}) and (\ref{aq11}).

\subsection{ On the Bethe ansatz }

In the limit where the harmonic attraction vanishes, the radial eigenfunction
(\ref{st3}) becomes a Bessel function $J_{\sqrt{\lambda}} (kr)$ and the 
relative energy is then $k^2/2$ where $k$ is a positive real. 
In comparison with a system of free particles, the asymptotic behaviour of 
the eigenfunction
\be{ba1}
J_{\sqrt{\lambda}} (kr) \sim_{r\to\infty} \sqrt{2\over\pi kr} \cos\Big(kr-
{\pi\over2}\sqrt{\lambda} -{\pi\over4}\Big)
\ee
sets up a phase shift proportional to the $\nu$-dependence of $\sqrt{\lambda}$. 
In the case of Bose and Fermi statistics, this phase shift can be viewed as a
sum of 2-body phase shifts in accordance with the Bethe hypothesis 
\cite{Marchioro}. In the case of mixed statistics, however, it is not the case 
and the Bethe hypothesis \cite{Yang} is no longer relevant.

\section{ A numerical evaluation for four bodies }

Leaving out the center-of-mass coordinate, I use the Jacobi coordinates 
\ba{ne1}
y &=& {1\over\sqrt{2}} (x_1-x_2), \nonumber\\
x &=& {1\over\sqrt{6}} (x_1+x_2-2x_3), \\
z &=& {1\over\sqrt{12}} (x_1+x_2+x_3-3x_4). \nonumber
\ea
The relative hamiltonian reads
\be{ne2}
\widetilde H=H_0 -\nu\left({\partial_y\over y}
+{\partial_x-\sqrt{2}\partial_z\over x-\sqrt{2}z} +{\rm c.p.}   
\right)
\ee
where c.p. stands for the cyclic permutations of $x_1,x_2,x_3$.
Expressing $x,y,z$ in terms of the spherical coordinates $r,\theta,\phi$, 
the wavefunction may be separated as $\widetilde\psi= \Xi(X) \widetilde R(r) 
\widetilde Y(\theta,\phi)$. The center-of-mass eigenstates are (\ref{st2})
and the radial eigenstates $R=r^{3\nu} \widetilde R$ are (\ref{st3}). 

The first angular eigenstates may be evaluated numerically by direct 
diagonalization of the matrix of the angular part of the hamiltonian 
in a basis of spherical harmonics $Y_{\ell,m}$. For simplification, 
the analysis is restricted to the symmetric wavefunctions under the 
exchanges of the particle coordinates $x_1,x_2,x_3$, that is, $Y_{\ell,m}$ 
have to be symmetrized under $\phi\to \phi+{2\over3}\pi$ and $\phi\to -\phi$.
In such a situation, there are only $4!/3!=4$ ground states.
Since the potential $(x-\sqrt{2}z)^{-1} (\partial_x-\sqrt{2}\partial_z)$
may be transformed into $z^{-1}\partial_z$ by means of a simple rotation,
one has to calculate the matrix elements of $x^{-1}\partial_x$ and 
$z^{-1}\partial_z$. In each case, one has to effect a simple Cauchy 
integration over $\phi$ and then the integration of a certain polynomial
by means of the Gauss-Legendre quadrature. Fig.2 displays the discretized 
values of $\sqrt{\lambda+{1\over4}}$ obtained by the numerical diagonalization 
of a $222\times222$ matrix for different values of $\nu$. One distinguishes 
the 4 ground states and some excitements. The Bose groundstate displays a
linear energy of slope 6. The other energies are not linear but their slope
is always 2 at $\nu=0$ and $\nu=1$.

\section{ Statistical mechanics in the thermodynamic limit }

In this section, the harmonic well has to be understood as a long distance
regulator alternative to the box \cite{comment}. For instance, expressing 
the nth cluster coefficient $b_n= Z_n-Z_{n-1}Z_1 +\ldots {1\over n}Z_1^n$ 
in terms of the $N$-body partition functions $Z_N$ \cite{Huang}, the 
$\omega\to0$ limit can be identified to the infinite box thermodynamic limit 
if the divergent factor $n^{-1/2} (\beta\omega)^{-1}$ surviving in $b_n$ is 
identified to an additive factor, namely $V\lambda_T^{-1}$ where $V$ is the 
box volume and $\lambda_T= \sqrt{ 2\pi/mk_BT}$ is the thermal wavelength.

In the case of Bose and Fermi statistics, the thermodynamic limit of the 
thermodynamical potential has been obtained by various means from the 
many-body spectrum in a harmonic well 
\cite{Sutherland,Dasnieres,Presutti,Murthy}. 
The result is in accordance with the thermodynamic Bethe ansatz 
and thus provides a confirmation of the Bethe hypothesis. Interesting enough, 
the particles obey Haldane's generalization of the Pauli principle 
\cite{Haldane}. Note that a correlation function has recently been 
analysed \cite{Ha,Forrester}.

In the case of Boltzmann statistics where no exchange symmetry is imposed 
to the wavefunction, I can calculate the cluster coefficients $b_1 =Z_1$, 
$b_2=Z_2 -{1\over2}Z_1^2$ and $b_3=Z_3 -Z_2Z_1 +{1\over3}Z_1^3$ from the 
Boltzmann partition functions $Z_N= {1\over N!}\,{\rm tr}\, e^{-\beta H_N}$. 
The thermodynamic limit gives $b_1= V\lambda_T^{-1}$, $b_2=0$ and 
the new result
\be{sm1}
b_3= {V\over\lambda_T} {1\over6\sqrt{3}} \left(
\Big( {3\over\pi}\arcsin {\cos\nu\pi\over2} \Big)^2 
-{1\over4} \right).
\ee
This last result is rather different from that of a system of particles
obeying Haldane statistics \cite{Haldane}.

\section{ On the multi-species Calogero model }

One considers the $N$-particle hamiltonian
\be{ms1}
H=\sum_{i=1}^N \big(-{1\over2}\partial_i^2 +{1\over2}\omega^2 x_i^2\big)
+\sum_{i,j<} {\nu_{ij}(\nu_{ij}-1)\over x_{ij}^2}
\ee
where the masses have been eliminated by a suitable redefinition on the 
coordinates and on the parameters. The discussion displayed between 
(\ref{m1}) and (\ref{st4}) is immediately generalizable to the multi-species
model if the change $\nu\to \nu_{ij}$ is done. However the wavefunction
is no longer in an irreducible representation of $S_N$. Moreover, the 
non-unitary transformation 
\be{ms2}
\psi= \prod_{i,j<} |x_{ij}|^{\nu_{ij}}\; \widetilde\psi
\ee
introduces three-body terms,
\be{ms3}
\widetilde H=\sum_{i=1}^N \big(-{1\over2}\partial_i^2 +{1\over2}\omega^2 
x_i^2 \big) -\sum_{i,j<}{\nu_{ij}\over x_{ij}}(\partial_i-\partial_j)
-{1\over2}\sum_{i,j,k\ne}{\nu_{ij}\nu_{ik}\over x_{ij}x_{ik}}.
\ee
By symmetrizing under the cyclic exchanges of $i,j,k$ and reducing to the
same denominator the term of the sum, one verifies that the three-body 
potential vanishes when $\nu_{ij}=\nu$.

The possibility to generalize the algebra of Raising and lowering operators 
to the multi-species Calogero model by the change $\nu\to\nu_{ij}$ 
in (\ref{so1})
have been studied in \cite{Furtlehener}. It appears that this new algebra 
of operators does not provided physical excitements except for center-of-mass 
and radial excitements. Nevertheless, every excitement provided is good 
at first perturbative order in $\nu_{ij}$.

Let me exemplify the multi-species Calogero model by a numerical and 
perturbative analysis of the 3-body problem. Only the angular part of the
eigenvalue problem have to be considered again,
\be{ms4}
\Lambda= -\partial_\phi^2 +{\nu_{12}(\nu_{12}-1)\over\sin^2\phi}
+{\nu_{23}(\nu_{23}-1)\over\sin^2(\phi+{2\pi\over3})}
+{\nu_{31}(\nu_{31}-1)\over\sin^2(\phi+{4\pi\over3})}.
\ee
The non-unitary transformation
\be{ms5}
\Phi(\phi)= \Big|\sin\phi\Big|^{\nu_{12}} 
\Big|\sin(\phi+{{2\pi}\over{3}}) \Big|^{\nu_{23}} 
\Big|\sin(\phi+{{4\pi}\over{3}}) \Big|^{\nu_{31}} 
\; \widetilde\Phi(\phi)
\ee
gives an operator
\ba{ms6}
\widetilde\Lambda &=& -\partial_\phi^2 +(\nu_{12}+\nu_{23}+\nu_{31})^2 
\nonumber \\
& & -2\nu_{12} \cot{\phi} \, \partial_\phi 
-2\nu_{23} \cot{(\phi+{2\pi\over3})} \, \partial_\phi \nonumber \\
& & -2\nu_{31} \cot{(\phi+{4\pi\over3})} \, \partial_\phi \\
& & +{\nu_{23}\nu_{31} \over \sin(\phi+{2\pi\over3}) \sin(\phi+{4\pi\over3})}
+{\nu_{31}\nu_{12} \over \sin(\phi+{4\pi\over3}) \sin\phi} \nonumber \\
& & +{\nu_{12}\nu_{23} \over \sin\phi \sin(\phi+{2\pi\over3}) } \nonumber
\ea
the matrix elements of which are well defined with the principal value 
regularization. Indeed, the angular wave function is periodic of period 
$2\pi$ and thus it may be expanded in Fourier series,
\be{ms7}
\widetilde\Phi(\phi)= {1\over\sqrt{2\pi}} \sum_{\ell=-\infty}^\infty c_\ell
e^{\ii \ell\phi},
\ee
and the matrix elements of $\widetilde\Lambda$ are expressed in the Fourier 
space as simple Cauchy integrals. The final result reads
\ba{ms8}
\langle \ell| \widetilde\Lambda |\ell'\rangle &=&
\ell^2\delta_{\ell=\ell'} +(\nu_{12}+\nu_{23}+\nu_{31})^2 \, 
\delta_{\ell=\ell'} \nonumber \\
& & +2\ell'\sign(\ell'-\ell) \, \delta_{\ell-\ell' \, {\rm even}} \nonumber \\
& & \times\left(\nu_{12} +\nu_{23}\, \jj^{\ell-\ell'} +\nu_{31}\, \jj^{2\ell
-2\ell'} \right) \\
& & +{\rm fact}(\ell-\ell') \, \delta_{\ell-\ell' \, {\rm even}} \nonumber \\
& & \times\left(\nu_{23}\nu_{31} +\nu_{31}\nu_{12}\, \jj^{\ell-\ell'} 
+\nu_{12}\nu_{23}\, \jj^{2\ell-2\ell'} \right) \nonumber
\ea
with $\sign(0)=0$ and ${\rm fact}(\ell-\ell')=0,-2,2$ respectively if 
$|\ell-\ell'|=0,2,4$ modulo 6. Diagonalizing a $361\times361$ matrix, 
I have obtained a numerical evaluation of the first eigenvalues $\lambda$. 
Note that the energy spectrum is built from the square root of $\lambda$ 
according to (\ref{st4}).

In the case $\nu_{ij}=\nu$, the numerical results reproduce the 
exact spectrum of Fig.1 as it should.

In the case $\nu_{12}=0$ and $\nu_{23} =\nu_{31} =\nu$, the first 
numerical values of $\sqrt{\lambda}$ are displayed in Fig.3. The spectrum 
is nearly linear in the quantum number $\ell$. However, the computation 
at second perturbative order around $\nu=0$ bears out the presence of 
some irregularities. Since the problem is invariant under $x_1\leftrightarrow 
x_2$, the non-degenerate perturbative theory may separately be used for 
the unperturbed bases $\{\cos(\ell\phi)\}$ and $\{\sin(\ell\phi)\}$. 
For instance, starting from the unperturbed eigenstate $\cos(\ell\phi)$ 
with $\ell=1$ modulo 3, one finds 
\be{ms9}
\sqrt{\lambda}= \ell-\nu +({\pi\over2\sqrt{3}} -{1\over\ell})\nu^2 +O(\nu^3)
\ee
Since these energies are not linear in the quantum number, there is no 
algebra of Raising and lowering operators in the multi-species Calogero model.

\section{ Conclusion }

The nature of the wavefunction of the Calogero model has been discussed
at length.

As regards the case of Boltzmann statistics, we have revisited the 
Raising and lowering operators and we have obtained some new 
exact eigenfunctions. One would like to derive the complete solution 
of this many-body problem, if possible analytically.

In the multi-species case, we have shown the impossibility of an algebra
of Raising and lowering operators. However, the question of the 
integrability of this model remains open.

At least, it would be desirable to have a better understanding of the 
commutation rules between the Raising and lowering operators 
of the one-species model.


\end{document}